\documentstyle[11pt,moriond,epsfig]{article}

\def\be{\begin{equation}}
\def\ee{\end{equation}}
\def\bea{\begin{eqnarray}}
\def\eea{\end{eqnarray}}

\def\Tr{\mbox{Tr}}
\def\S{\mbox{S}}
\def\D{\mbox{D}}
\def\I{\mbox{I}}
\def\M{\mbox{M}}

\def\max{\mbox{max}}
\def\V{\mbox{V}}
\begin{document}
\vspace*{4cm}
\title{The Effect of Interaction  on Shot Noise in The Quantum Limit}

\author{D.B. Gutman and
Yuval Gefen}

\address{Department of Condensed Matter Physics,
 The Weizmann Institute of Science 76100, Rehovot, Israel}
\maketitle\abstracts{
We calculate the  current and the current noise in
 a diffusive micro-bridge
 in the presence of electron-electron interactions.
 Out of equilibrium
 the fluctuation-dissipation theorem (FDT) does not apply, hence
 these two quantities are not simply related to each other.
 For a two-dimensional electron gas (2DEG) we   obtain
 logarithmic singularities in the low frequency limit.
 PACS~Nos.~71.10.Ay, 71.23.An, 73.50.Td}
\section{Introduction}
The physics of  non-equilibrium mesoscopic systems  has been the subject of
 extensive theoretical and experimental
 research for more than a  decade
 \cite{Kogan_book,Blanter,Beenaker_proc}.
 The first step \cite{Landauer_priv_com,Khlus,Lesovik,Buttiker}
 was the understanding that Fermi statistics has
 a dramatic effect on the noise in the  quantum limit. Employing various
 methods ( e.g.  semiclassical approximations
 \cite{Khlus} the Landauer scattering states approach
 \cite{Khlus,Lesovik,Buttiker} and diagrammatic Keldysh techniques \cite{Altshuler-Levitov})  it has been recognized that the
 zero frequency \cite{1/f}
 zero temperature noise vanishes in the limit of perfect transparency, while
 the conductance remains finite. For a multichannel geometry one obtains
 \begin{eqnarray}
\label{u1}
 S(0)=\frac{e^2}{\pi \hbar}\sum_n T_n(1-T_n)eV \,\, ,
 \end{eqnarray}
 where $\{T_n\}$ are the channel transparencies and $V$ is the  voltage
 drop across the junction under consideration.
 This result may  be used when discussing disordered conductors. Upon averaging
 over  an ensemble of coherent diffusive systems
 (the size of which  is smaller than any of the  inelastic, dephasing and
 localization lengths),  one obtains  a universal $1/3$ reduction  from
 the  traditional  expression for shot-noise \cite{Beenakker_Buttiker},
 $S=e <I>$
( the expectation value with respect to the effective action,
denoted by  $< >$ , includes configuration   averaging).

 While the physics of non-interacting electrons ( especially when
 disorder averaging is involved) turns out to be quite universal, this is
 not the case  when electron-electron interactions are present.
 Nevertheless, when the dynamics of the electrons is classical, rendering
 quantum mechanical  interference  effects negligible, one may resort to the
 physically appealing kinetic equation technique, or more specifically,
 employ the  Kogan-Shulman \cite{Shulman}  approach.
 Under such conditions  it is possible to  define an effective
 single-particle distribution function which fluctuates in time and space.
 The characteristics of this distribution function , which affect the  noise,
 depend on  non-universal factors such as the  external screening
\cite{Nagaev2,Nagaev3}, the geometrical details of the contacts
 \cite{Likharev} and the various  inelastic processes rates
 \cite{Blanter,Nagaev95} .
 One may then  consider two limiting cases, namely the
 short  and the  long  energy relaxation length.
 Correspondingly, and depending
 on the frequency range , one may obtain different suppression factors
 ( see e.g.\cite{Nagaev95,Henny}), or even a noise spectrum which depends on
 the spatial coordinate
 \cite{Nagaev2}.
 The   dependence of the shot noise on both  Fermi statistics and
 on the value of the discrete elementary charge
 has become manifest through experiments on  fractional quantum Hall systems, where electron-electron interaction redefine the quasi-particle charge
\cite{Resnikov}.

 Apart from the later effect, electron-electron interactions
 may introduce further  non-trivial signatures.
 For interacting systems {\it at equilibrium} this has been studied  by
 Altshuler and Aronov  \cite{Altshuler}, who found small albeit  singular
 ( in the temperature) corrections to
 the linear conductivity ( hence, by FDT, to the equilibrium noise). These
 corrections to the noise are the result of  an  interplay
 between disorder and
 interaction, and depend on the coherence of the electrons.
 One might expect an analogous type of physics  out of equilibrium.
 Pursuing the study of the effect of
 interactions {\it beyond} the scope of the kinetic equation and well
 into the
 quantum regime  is the main focus of the present work.
\section{Results}
 We consider here a  high mobility  disordered metallic  bridge,  whose
 {\it Ohmic} conductance is  $G_{Ohm}$. This bridge is connected to
 two ideal leads , each merging
 adiabatically  onto a respective reservoir. The  reservoirs are assumed
 to be at
 equilibrium with chemical potentials differing by $eV$.
 The linear dimensions of the bridge  are all
 larger than the  elastic
 mean free path  , $l$, yet  much shorter than  the  inelastic length,
 $l_{in}$.
As the problem is technically challenging, we resort to a model
calculation ,taking the interaction to be short-ranged.
 Considering a low electron-electron collision rate, one may assume
 the temperature $T$ to be constant throughout the
 system \cite{Equilibrium}.
 It is now convenient to define $\zeta=\max\{eV,T\}$; $g_{\Box}$  will denote
 the  dimensionless conductance per square, $\hbar \nu D$, where $\nu$
 is the single electron density-of-states and  $D$  is the diffusivity.
 Our main result concerns the zero frequency noise which  is written as
\begin{eqnarray}&&
\S= \S_0 +\delta \S \, \, ,
\end{eqnarray}
 where the interaction
 corrections to the zero frequency noise are given by
\begin{eqnarray}&&
\label{e29} \delta \S\!=\!\frac{G_{Ohm}}{6\pi^2g}
\bigg[2T\ln\left(\!\frac{\zeta\tau}{\hbar}\!\right)
\!+\!eV\!\coth\left(\!\frac{eV}{2T}\!\right)\!\!
\bigg(\!\ln\left(\!\frac{\zeta\tau}{\hbar}\!\right)\!+\!\ln\left(\!\frac{T\tau}{\hbar}\!\right)\!\!\bigg)\!\bigg].
\end{eqnarray}
  This result should be compared with  the  corresponding
 correction to the conductance. We  have obtained
 \begin{eqnarray}&&
 \label{a60}
 \delta G_{Ohm}=\frac{G_{Ohm}}{2\pi^2 g}\ln\left(\frac{\zeta\tau}{\hbar}\right) \,\, .
 \end{eqnarray}

 At equilibrium , as long as FDT is applicable, the interaction  correction
 to the current noise may be cast  as a correction  to the
 Ohmic part of the conductance \cite{Altshuler} .
 Our results , Eqs.(\ref{e29}) and (\ref{a60}),
 demonstrate  that  out of equilibrium this is not  any more the case:
 while corrections to the  conductance are determined by
 the larger of  the temperature and the applied voltage,
 the  most singular correction to the noise involves a
 temperature-dominated  logarithmic singularity.
\section{The Approach}
 We now present a brief description of our derivation.
Full account of the analysis is presented  elsewhre\cite{GG}
 We took notice of the work of Kamenev and Andreev
\cite{Kamenev} , who had  addressed
 the physics of disordered interacting electron gas by deriving a sigma-model
 description \cite{Fin} defined
 on a Keldysh contour ( see also Ref. \cite{Chamon}). In their work
 \cite{Kamenev} the
 interaction has been  accounted for  by expanding around an (approximate)
 interaction-dominated  saddle point , obtained through  gauge
 transformation.
 Here we pursue  the Kamenev-Andreev approach further, to include
 non-equilibrium  effects \cite{GG}.

 Our starting point is  the Lagrangian density
 \begin{eqnarray}
 \hat{{\cal L}}=\overline\Psi[\hat{G}_0^{-1}-U_{dis}]\Psi -\overline\Psi \overline\Psi \hat{\V}\Psi\Psi
 \,\, .
 \end{eqnarray}

 Here $\hat{V}$ represents the  electron-electron interaction ;
 the Schr\"odinger operator of an electron of mass $m$ in the
 presence of a  vector potential $a$ is given by
 \begin{eqnarray}
 \hat{G}^{-1}_0=i\hbar\frac{\partial}{\partial t}+\frac{\hbar^2 }{2 m}\left(\nabla-a\right)^2+\mu\,\, .
 \end{eqnarray}
 All energies are measured from the chemical potential $\mu$.
 The action now contains an integral over space and an integral in time
 over the   Keldysh contour .
 Such a representation is  particularly  convenient for averaging over
 disorder, as
 the generating  functional $Z[a]$
 is identically  equal to $1$ when the external sources on the forward
 and backward paths are equal.

 Next , one  averages   over  the
 $\delta$-correlated disorder. We then  perform   a Hubbard-Stratonovich
 transformation, introducing  the bosonic fields $Q$ and $\Phi$
 which decouple
 the non-local-in-time term (generated  by disorder averaging)  and the
 non-local-in-space term ( produced by the Coulomb interaction ) respectively.
 In terms  of the bosonic matrix field  $Q$ and $ \Phi$ the action now reads
 \begin{eqnarray}&&
iS[\hat Q,\hat\Phi]=i\Tr\{\Phi^T \V^{-1}\gamma_2\Phi\}\!
-\frac{\pi\nu}{4\tau} \Tr\{\hat Q^2\} \! +  \nonumber \\&&
\Tr\ln \Big[
\hat G_0^{-1} \! +\!  \frac{i\hat Q \sigma_3 }{2\tau}
\! +\! \hat \phi_{\alpha}\hat\gamma^{\alpha}
\Big].
\end{eqnarray}

 Unlike  the  original treatment of   Ref.\cite{Kamenev}, we
 presently  try to avoid the overwhelming task of finding an
 interaction-gauged saddle-point solution  out of equilibrium . Instead, we
 consider here  short-ranged interaction which , we believe, captures
 the essentials of the problem at hand:
 \begin{eqnarray}&&
 \hat{\V}(x-x')=\Gamma \delta(x-x') \, .
 \end{eqnarray}
 Hereafter $\Gamma$ is  assumed to be small $(\Gamma\nu \ll 1)$.
 For  weak disorder  $(\epsilon_F \tau \gg 1$ ,$\tau$ being
 the elastic mean free
 time) it is possible to separate the slow
 and fast degrees of freedom . Expanding around the {\it noninteracting}
 saddle point yields  the
 following effective action
 \begin{eqnarray}&&
 \label{e52}
 iS=\!i\Gamma^{-1}\Tr\{\Phi^T\gamma_2\Phi\}\!-\!\frac{\pi\nu}{4}\Tr\{D(\nabla\!Q)^2\!\!-\!4i\!(\phi_\alpha\gamma^\alpha\!+\!\hat{\epsilon})Q\}\,\,,
 \end{eqnarray}
 with the usual non-linear constraint
 $Q^2=1$. Here  $\Phi^T=(\phi_1,\phi_2)$  and
 \begin{eqnarray}
 \gamma_1=\left(\matrix {1 & 0\cr 0 & 1 \cr}\right) , \;
 \gamma_2=\left(\matrix {0 & 1\cr 1 & 0 \cr}\right).
 \end{eqnarray}
 Within this model one may calculate the correlation function
 of the fluctuations of the electric potential
 \begin{eqnarray}&&
 \label{cor_phi}
 \langle \phi_i(r,\omega)\phi_j(r',-\omega)\rangle\! =\!-i \Gamma \delta(r-r')\gamma_2^{i,j}.
 \end{eqnarray}
 Let us first consider the non-interacting scenario , stressing a few points
 that
 were implicit in previous works \cite{Nagaev1,Altshuler-Levitov}.
  The  gapless fluctuations of $Q$  around the saddle point
 are conveniently parameterized as
 \begin{eqnarray}&&
 \label{e24}
 Q=\Lambda\exp\left(W\right) \nonumber\,\, , \\&&
 \Lambda W+W \Lambda=0 \,\, .
 \end{eqnarray}
 To satisfy   the non-linear constraint on $Q$  and , at the same time ,
 conform to  Eq. (\ref{e24}),
 one may use the representation
 \begin{eqnarray}
 W_{x,\epsilon,\epsilon'}=\left(
 \matrix {F_{x,\epsilon}
 \bar{w}_{x,\epsilon,\epsilon'} &-w_{x,\epsilon,\epsilon'}+F_{x,\epsilon}\bar{w}_{x,\epsilon,\epsilon'}F_{x,\epsilon'}\cr-\bar{w}_{x,\epsilon,\epsilon'}  & - \bar{w}_{x,\epsilon,\epsilon'}F_{x,\epsilon'}\cr}
 \right) \,\,   ,
 \end{eqnarray}
 where the fields $\omega, \bar{\omega}$ are unconstrained.
 Expanding the action  to second order  in the soft modes
 yields for  the free part of the action
 \begin{eqnarray}
 iS[W]=iS^0[W]+iS^1[W]+iS^2[W] \,\, .
 \end{eqnarray}
 The zeroth order term vanishes at the saddle point.
 The  linear part in $W$  should vanish as well, yielding an equation for
 $F$  which, together with the appropriate boundary conditions, determines
 $F$ uniquely
 The quadratic part of the action is equal to
 \begin{eqnarray}&&
 \label{e55}
 iS^2[W]=\frac{\pi\nu}{2}
 \bigg[\bar{w}_{x,\epsilon,\epsilon'}
 [-D\nabla^2+i(\epsilon-\epsilon')]w_{x,\epsilon',\epsilon}
 -\nonumber \\&&
 D \nabla F_{x,\epsilon}\bar{w}_{x,\epsilon,\epsilon'}\nabla F_{x,\epsilon'}
 \bar{w}_{x,\epsilon',\epsilon}
 \bigg]\,\,.
 \end{eqnarray}
 The last term in (\ref{e55})
 is
 responsible for equilibration along the longitudinal coordinate; it is
 absent at equilibrium.

 Turning our attention now to the diffusive constriction  geometry,
 we specifically consider the  slow
 energy relaxation time  limit. The
 distribution function is then  readily calculated \cite{Nagaev1,Kamenev}.
 One may next evaluate  various  correlation functions
 for this   system out of equilibrium. For  our
 unconstrained Gaussian action,
 Eq. (\ref{e55}),  they can be  written in terms of the
 diffusion propogator \cite{diff_propogator}
 \begin{eqnarray}&&
 \label{y1}
 \langle w(x,\epsilon_1,\epsilon_2) \bar{w}(x',\epsilon_3,\epsilon_4)
 \rangle=2\delta_{\epsilon_1,\epsilon_4}
 \delta_{\epsilon_2,\epsilon_3} \D(x,x',\epsilon_1-\epsilon_2)
 \nonumber \\&&
 \langle w(x,\epsilon_1,\epsilon_2)w(x',\epsilon_3,\epsilon_3)\rangle=-\delta_{\epsilon_1,\epsilon_4}
 \delta_{\epsilon_2,\epsilon_3}
 2\pi \nu D \nonumber \\&&
 \D_{\epsilon_1-\epsilon_2,x,x_1}\nabla F_{\epsilon_2,x_1} \otimes
 \nabla F_{\epsilon_1,x_1} \D_{\epsilon_2-\epsilon_1,x_1,x'} \,\, .
 \end{eqnarray}
 Employing the above building blocks we may calculate the symmetrized
 current-current correlation function
 \begin{eqnarray}
 \S(t,x,t',x')=\frac{1}{2}\big[
 \langle \langle \hat{I}(t,x)\hat{I}(t',x') + \hat{I}(t',x')\hat{I}(t,x)
 \rangle\rangle\big] \,\, .
 \end{eqnarray}
 Under stationary conditions  it is a function of the difference of its
 arguments.
 Relying on the identity $Z[a_{+},a_{-}=0]=1$ ( where the indices  refer
 to the symmetric (``classical'') and the antisymmetric (``quantal'') combinations
 over the Keldysh  branches) ,
 we express $\S$ as
 \begin{eqnarray}
 \S(x,t;x',t')=-\frac{e^2}{4}\frac{\delta^2 Z[a]}{\delta a_2(x,t)\delta a_2(x',t')}
 \,\, .
 \end{eqnarray}
 After functional differentiation  one obtains the following equation
\begin{eqnarray}&&
\label{e7}
\S_{x,t;x',t'}\!=\!
\frac{e^2\pi\nu\!D}{4}\bigg<\!\I^D_{x,t,t'}\!\delta_{x,x'}\!-\!
\frac{\pi\nu\!D}{2}\M_{x,t}\!\M_{x',t'}\!\bigg>\!-\!\langle\I\rangle^2,
\end{eqnarray}
where we have introduced the notation
\begin{eqnarray}&&
\I^D_{x,t}=\Tr\!\bigg\{\!Q_{x,t,t'}\gamma_2Q_{x',t',t}\gamma_2\!-
\!Q_{x,t,t'}Q_{x',t',t}\bigg\}\delta_{x,x'} \nonumber \\&&
\M_{x,t}=\Tr\bigg\{\big(Q_{x,t,t_1}\nabla Q_{x,t_1,t}-
(\nabla Q_{x,t,t_1})Q_{x,t_1,t}\big)\gamma_2 \bigg\} \,\, .
\end{eqnarray}
Expanding Eq. (\ref{e7}) to first  order in $1/g$ and to zeroth
order in $\Gamma$ , we recover the results of the direct diagrammatic
analysis of this problem \cite{Altshuler-Levitov}.
The extra power of $g$ in the  second term
( a non-local part ) cancels with $\langle\,\I\,\rangle^2$.
 At high frequencies, dynamical fluctuations of the local  density are
 feasible, leading to  the dependence of the noise on the the
spatial coordinates;
 this is not anymore the case  at low frequencies,
 $\omega \ll E_{Th}$ , which is a direct
 consequence of particle conservation.
 In this limit $\S$ is  given as a  sum of three
 equilibrium spectral functions \cite{Zero}
 \begin{eqnarray}&&
 \label{e25}
 \S_0(\omega)=\frac{1}{6}[4\S^{eq}(\omega)+\S^{eq}(\omega+eV)+\S^{eq}(\omega-eV)]\,\, ,
 \end{eqnarray}
as has been found earlier \cite{Khlus}.
\section{The Effect Of e-e Interaction}
So far we have considered the leading part of the current noise.
Now on we discuss corrections due to  e-e interactions.
We note that under the conditions specified above,
the non-equilibrium quasi-particle distribution function
is essentially of a (temperature and interaction smeared) double-step form.
We first evaluate  the interaction
correction to the stationary value of electron current, $\delta\I^{ee}$.
Employing the relation
 \begin{eqnarray}&&
 \label{f1}
 I=\frac{-e }{2i}\frac{\delta Z[a]}{\delta a_2} \,\, ,
 \end{eqnarray}
 and using the action (Eq. \ref{e52}),  we obtain
 \begin{eqnarray}&&
 \label{f2}
 \delta \I^{ee}=\frac{e\pi \nu D (i\pi \nu)^2}{4}
 \langle \M \, \Tr\{\phi_\alpha\gamma^\alpha Q\} \Tr\{\phi_\beta \gamma^\beta Q \}\rangle_{0}\,\,.
 \end{eqnarray}
Employing the rules of Eq.\ref{y1} , this leads to
\begin{eqnarray}&&
 \label{z28}
\delta \I^{ee}= \langle \I \rangle \frac{\Gamma\nu }{2(2\pi)^2g} \ln\left(\frac{\zeta\tau}{\hbar}\right)  \,\, .
\end{eqnarray}
Let us turn now to the calculation of the interaction
correction to the current noise. To this end one evaluates
  Eq. (\ref{e7}),
employing  the interaction dependent  action, Eq. (\ref{e25}) .
In   first order in the interaction we obtain
\begin{eqnarray}&&
\label{z34}
\delta\S^{ee}(0)=\frac{e^2\pi\nu D}{4}
\bigg<[I^D_{x,t,t'}\delta_{x,x'}\!-\!\frac{\pi\nu D}{2}M_{x,t} M_{x',t'}]\nonumber \\&&\Tr\{\phi_\alpha\gamma^\alpha Q\}\Tr\{\phi_\beta \gamma^\beta Q \}\bigg> \,\, .
\end{eqnarray}
To proceed , we first  carry  out the averaging  over the fields
 $\phi$ in   Eq. (\ref{z34})  employing  Eq. (\ref{cor_phi}) .
This leads to a $\Gamma$ dependent  expression written solely  in terms
of the  matrices $Q$.
We then expand  in the fields $\omega, \bar{\omega}$ and
perform  contractions according to the rules of Eq.(\ref{y1}).
For  frequencies smaller than the Thouless frequency ,
the current-current correlation function does not depend
on the choice of the cross-section : integration over the
spatial coordinates will lead to the suppression of space-derivative
terms.
We finally obtain
\begin{eqnarray}&&
\label{e30}
\delta\!S^{ee}\!\!=\!\!\frac{G\!\Gamma\nu}{12\!\pi^2g}\!\!\bigg[\!2T\!\ln\left(\!\frac{\zeta\tau}{\hbar}\!\right)
\!\!+\!eV\!\!\coth\left(\!\frac{eV}{2\!T}\!\right)\!\!\bigg(
\!\!\ln\left(\!\frac{\zeta\tau}{\hbar}\!\!\right)\!+\!\ln\left(\!\!\frac{T\tau}{\hbar}\!\!\right)\!\!\bigg)\!\bigg].
\end{eqnarray}
The derivation of Eqs. (\ref{z28}) and (\ref{e30}) involves some
controlled approximations. The use of a short range interaction
(assumed to be small,  $\Gamma \nu \ll 1$) renders them model
dependent.

To establish a relation with the ``real''  screened Coulomb interaction, we
further propose the following heuristic picture:
we argue that the   correct prefactor in the expression for
 $\delta \I^{ee}$   should  not depend
 strongly on either the temperature or the  voltage ; hence it  can be
 restored based on  equilibrium calculations \cite{Altshuler} ,
 \begin{eqnarray}&&
 \label{f8}
 \delta \I^{ee}= \langle \I \rangle \frac{1}{2\pi^2 g} \ln\left(\frac{\zeta\tau}{\hbar}\right)  \,\, .
 \end{eqnarray}
 Comparing Eqs.(\ref{z28})    and (\ref{f8}) we are
 able to express $\Gamma$ in terms of the model-independent parameters of
 the problem , which leads to Eq.(\ref{a60}).  This mapping will
 be employed for the expression for  the noise
 as well, leading to Eq. \ref{e29}.
Hereafter we have assumed  $\zeta \gg E_{Th}$, which is readily
satisfied for macroscopic samples.
We note that both the voltage and the temperature may serve as
cut-offs for  the logarithmic singularity.
For $T \gg eV$ the singularity is governed by $T$,
resulting in the Altshuler-Aronov correction to the conductance.

Considering the asymptotic high temperature behavior of Eq.(\ref{e29}),
one notes  that the equilibrium part is dominant.
By the FDT Eq. (\ref{e29}) is related  to the Altshuler-Aronov
correction to the conductance \cite{Altshuler} , which is obtained
from Eq. (\ref{a60})  by substituting $\zeta=T$:
\begin{eqnarray}&&
\delta\S=\frac{G_{Ohm}}{\pi^2g}T\ln\left(\frac{T\tau}{\hbar}\right),
\end{eqnarray}
or, alternatively, it is related to the correction to the
mean current, Eq.(\ref{f8}).

In the large voltage limit, current fluctuations are
dominated by the bias dependent shot noise. Note, though,
that the  interaction corrections to it are still determined by the
{\it temperature},
\begin{eqnarray}&&
\label{r2} \delta\S=\frac{G_{Ohm}}{6\pi^2 g}
e|V|\ln\left(\frac{T\tau}{\hbar}\right) \,\, .
\end{eqnarray}
The  FDT  breaks down manifestly ,
and the results for the current (Eq. \ref{f8}) and for the current noise
(Eq. \ref{r2}) are no longer  related to each other in a simple way.
Moreover,  the interaction correction to the current noise
is larger
(for large bias ) than the corresponding  correction to the mean current.
On a qualitative level we note that
within a   classical picture,the  electron-electron interaction
may indeed lead to strong suppression of current fluctuations
without affecting the mean current much,
in agreement with our quantitative analysis.
In other words, the  relation  between the  mean current
and the  noise , $S \sim eI$, which is
valid for Poisson-like processes, is no longer applicable.
While we have presented here specific expressions
for the interaction corrections
in two dimensions, we do expect
effects of similar nature to take place at
other dimensions as well.

\begin{figure}[t]
\rule{5cm}{0.2mm}\hfill\rule{5cm}{0.2mm}
\vskip 9.5cm
\includegraphics{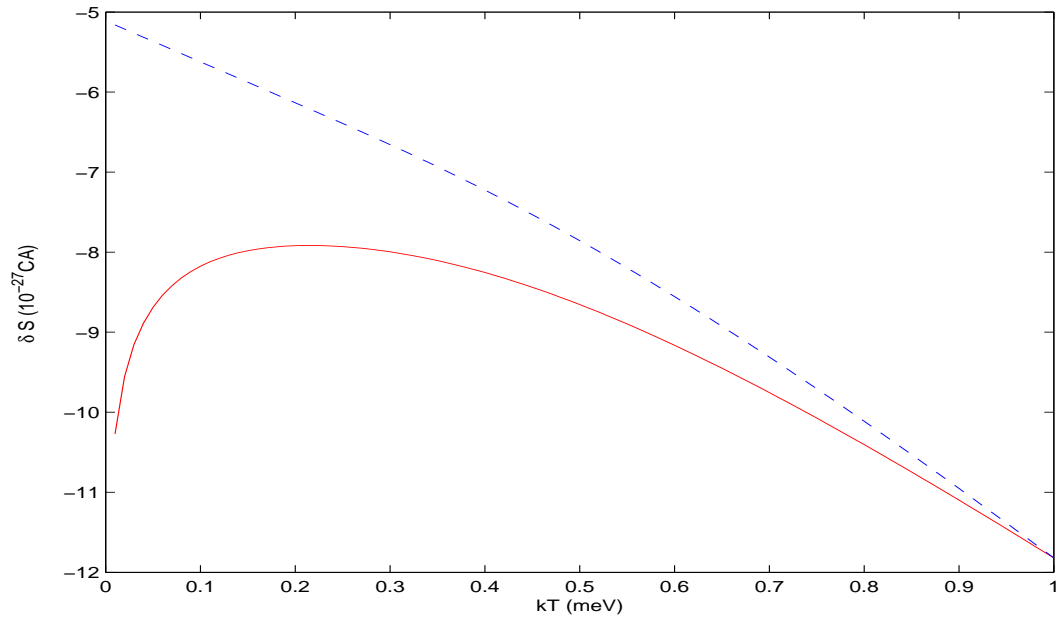}
\rule{5cm}{0.2mm}\hfill\rule{5cm}{0.2mm} \caption{Interaction
correction to the noise (solid line) and the "naive  expectation"
(dashed line), see text for definitions.}
\end{figure}
\section{Experimental Relevance}
>From the experimental point of view and for standard range of
parameters, the effect we predict here is quite small (i.e., the
logarithm of the ratio between the voltage and the temperature
dominated corrections is typically close to unity. However, the
incipient logarithmic singularity found here may,  in principle,
be spotted by carefully scanning the temperature dependence of the
noise. Moreover, one can push the  temperature and the voltage to
values where the effect is particularly enhanced. Fig[1] presents
a comparison of the corrections to the zero frequency current
noise derived here (solid line) with a ``naive'' prediction
(dashed line). The latter refers to using a standard expression
for the shot noise, where we insert the non-equilibrium
interaction corrections for $G_{Ohm}$. Here our two-dimensional
metallic film is assumed to have a square geometry. Its
conductance measured in units of quantum conductance  is taken to
be $G_{Ohm} =10 \frac{e^2}{\hbar}$, and the value of the elastic
mean free time is chosen to be  $\tau =10^-2\frac{1}{meV}$. The
applied bias $V=1meV$ and the temperature varies in the range
$0.001{\rm meV} < kT< 1{\rm meV}$. As the temperature decreases,
the discrepancy between the ''naive expectation" and the correct
result becomes increasingly pronounced.

\section*{Acknowledgments}
We acknowledge discussions with
A.M.~Finkelstein, A.~Kamenev, D.E.~Khmel'nitskii, L.S.~Levitov, A.D.~Mirlin,
M.~Rokni, B.Z.~Spivak and useful input on expermintal
aspects from D.~Prober, M.~Reznikov and R.~Schoelkopf. Y.G. acknowledges the
hospitality of and the interaction with B.L.~Altshuler at Princeton/NEC.
This work was
supported by the U.S.-Israel Binational Science Foundation,
the DIP Foundation, the Israel Academy of Sciences
and Humanities-Centers of Excellence Program,
and by the German-Israeli Foundation (GIF).


\begin{thebibliography}{99}
\bibitem{1/f} One should also keep in mind that this type of analysis is applicable
only in the case when shot or Nyquist noise are dominant,
and it is justified to neglect  $1/f$ noise,
i.e. the ``zero frequency'' is assumed to be finite.
\bibitem{Equilibrium} By this we do not imply that the system
is at equilibrium.On the contrary, the distribution
function  of the electron gas is a weighted sum of two
Fermi-Dirac functions, with different chemical potentials but the
same underlying $T$.
\bibitem{diff_propogator}
The diffusion propagator  $\D$ is defined through
$\left(-i\omega + D\nabla^2 \right)\D(\omega,r,r') =\frac{1}{\pi\nu}\delta(r-r') \,\, ,$
 with the appropriate boundary conditions.
\bibitem{diffusion_kineq}
In order for  the diffusion approximation for the kinetic  equation to be
valid ,  deviations  from equilibrium should be small, i.e.
the energy gain between two consecutive  elastic
 scattering is small in comparison with the Fermi energy ( $(eU/\epsilon_F) (l/L) \ll 1$ ) .
\end{thebibliography}
\end{document}